\documentclass[prb,twocolumn,superscriptaddress,showpacs,amsmath,amssymb]{revtex4-1}

\usepackage{amsmath}
\usepackage{amssymb}
\usepackage{graphics}
\usepackage{graphicx}
\usepackage{dcolumn}
\usepackage{txfonts}
\usepackage[pdfpagemode=UseNone,pdfstartview=FitH,colorlinks=true,linkcolor=blue,urlcolor=blue,anchorcolor=blue,citecolor=blue]{hyperref}
\allowdisplaybreaks[4]

\newcommand{\beq}{\begin{equation}}
\newcommand{\eeq}{\end{equation}}
\newcommand{\beqa}{\begin{eqnarray}}
\newcommand{\eeqa}{\end{eqnarray}}

\usepackage{color}
\usepackage{soul,xcolor}
\usepackage{ulem}
\setstcolor{red}

\begin{document}
\title{Control of the Two-Electron Exchange Interaction in a Nanowire Double Quantum Dot}
\author{Zhi-Hai Liu }
\affiliation{Interdisciplinary Center of Quantum Information and Department of Physics, Zhejiang University, Hangzhou 310027, China}
\affiliation{Quantum Physics and Quantum Information Division, Beijing Computational Science Research Center, Beijing 100193, China}

\author{O. Entin-Wohlman}
\email{oraentin@bgu.ac.il}
\affiliation{Raymond and Beverly Sackler School of Physics and Astronomy, Tel Aviv University, Tel Aviv 69978, Israel}
\affiliation{Physics Department, Ben Gurion University, Beer Sheva 84105, Israel}

\author{A. Aharony}
\affiliation{Raymond and Beverly Sackler School of Physics and Astronomy, Tel Aviv University, Tel Aviv 69978, Israel}
\affiliation{Physics Department, Ben Gurion University, Beer Sheva 84105, Israel}

\author{J. Q. You}
\email{jqyou@zju.edu.cn}
\affiliation{Interdisciplinary Center of Quantum Information and Department of Physics, Zhejiang University, Hangzhou 310027, China}
\affiliation{Quantum Physics and Quantum Information Division, Beijing Computational Science Research Center, Beijing 100193, China}

\begin{abstract}
The two-electron exchange coupling in a nanowire double quantum dot (DQD) is shown to possess Moriya's anisotropic superexchange interaction under the influence of both the Rashba and Dresselhaus spin-orbit couplings (SOCs) and a Zeeman field. We reveal the controllability of the anisotropic exchange interaction via tuning the SOC and the direction of the external magnetic field. The exchange interaction can be transformed into an isotropic Heisenberg interaction, but the uniform magnetic field becomes an effective inhomogeneous field whose measurable inhomogeneity reflects the SOC strength.  Moreover, the presence of the effective inhomogeneous field gives rise to an energy-level anticrossing in the low-energy spectrum of the DQD. By fitting the analytical expression for the energy gap to the experimental spectroscopic detections [S. Nadj-Perge {\it et al.}, {\color{blue} Phys. Rev. Lett. {\bf108}, 166801 (2012)}],  we obtain the complete features of the SOC in an InSb nanowire DQD.
\end{abstract}
\date{\today}
\maketitle

\section{Introduction}

Achieving an effective manipulation of the electron spins is of essential importance in the spin-based quantum information processing (see, e.g., Refs.~\onlinecite{Hanson2008,Vandersypen2007}).
It has been shown that the double quantum dot (DQD) is experimentally convenient for implementing logical gate operations.~\cite{Petta2005,Coppersmith2012,Koh2012,Bertrand2015}
In this case, the two-spin manipulation can be based on the exchange interaction in a DQD.~\cite{Taylor2007,Burkard1999} Generally, the intrinsic exchange interaction in a system results in a specific alignment of the spins. For example, the ferromagnetic exchange interaction leads to spin polarization,~\cite{Atodiresei2010} and the Dzyaloshinskii-Moriya (DM) exchange interaction induces the spin texture,~\cite{Moriya1960,Han2013,Rosch2013} which may give rise to skyrmion excitations in magnetic crystals.~\cite{Boni2009,Tokura2010}
Therefore, it is of great importance to realize the tunability of the exchange interaction between electrons.

Owing to the spin-orbit coupling (SOC), the spin degree of freedom is correlated with the orbital degree of freedom for electrons.~\cite{Kouwenhoven2010,You2013,Ban2012,Rashba1984,Levitov2003,Trif2008} In the absence of the SOC, the combined effects of the Coulomb interaction and the Pauli exclusion principle in the DQD give rise to the isotropic Heisenberg exchange interaction between electrons.~\cite{Taylor2007,Burkard1999} The presence of the SOC in semiconductor nanostructures mediates an anisotropic exchange interaction between electrons.~\cite{Kaplan1983,Baruffa2010,Starykh2008,You2014} However, the anisotropic exchange interaction can be mapped via an unitary transformation onto an isotropic Heisenberg interaction in the absence of an external magnetic field.~\cite{Kaplan1983,Kavokin2001,Shekhtman1993,Aharony1992,Imamura2004} Thus, the SOC seems only to have trivial influences on the exchange interaction. As shown here, this is not the case when an external magnetic field is present.

In the recent decade, the quasi-one-dimensional nanostructure with SOC has aroused much attention. Specifically, it can implement fast spin manipulations via an electric field.~\cite{Petersson2012,Sadreev2013,Berg2013,Nitta1997}  Also, it can act as
an effective spin splitter between two spin reservoirs~\cite{Shekhter2013,Shekhter2016} and offer a possible platform for searching the Majorana fermions in the superconductor-semiconductor hybrid systems.~\cite{Deng2016,Li2017,sau2010,Sau2012} Moreover, the SOC in a carbon nanotube plays an important role in determining spin transport properties,~\cite{Laird2015,Diniz2014} and facilitates unconventional superconductivity.~\cite{Hels2016}
In this paper, we investigate the two-electron exchange interaction in a symmetric nanowire DQD in the presence of a strong SOC and an external Zeeman field.
In the strong intradot Coulomb repulsion regime,~\cite{Wiel2003} the effective Hamiltonian describing the two electrons consists of a Zeeman term and a Moriya's anisotropic superexchange interaction.~\cite{Aharony1992,note00}
This anisotropic exchange interaction depends on the SOC strength in the material and can be manipulated by regulating the direction of the external magnetic field.
Furthermore, we show that when the anisotropic exchange interaction is transformed to an isotropic Heisenberg interaction, the uniform magnetic field becomes an effective inhomogeneous field, with the inhomogeneity depending on the SOC strength.

Under the effects of the SOC, there is an energy-level anticrossing, corresponding to the singlet-triplet splitting, in the low-energy part of the two-electron spectrum of the DQD,~\cite{Pfund2007,Perge2012,Fasth2007}
which is induced by the inhomogeneity of an effective magnetic field
in a symmetric nanowire DQD. More interestingly, based on the effective magnetic field we obtain an analytical expression for the singlet-triplet splitting.  By fitting the analytical formula to the experimental curve in Ref.~\onlinecite{Perge2012},  we extract the strength and direction of the SOC in an InSb nanowire DQD. The consistency between our theoretical results and the experimental analyses verifies our theory.
Moreover, our new results reveal that the spectroscopy measurements in the presence of an external Zeeman field can identify {\it separately} the Rashba and Dresselhaus parts of the SOC. For the existing experiments, the latter is very small for symmetry reasons. However, one can use other crystals, with other symmetry directions, and then both terms may appear.~\cite{Hofmann2017,Wang2018}
To implement quantum computing with large-SOC semiconductors, Bonesteel {\it et al.} demonstrated the elimination of the first-order SOC by tailoring the exchange coupling between two coupled spins,~\cite{Bonesteel2001} and then they exploited the passive switching of the exchange coupling to construct quantum logic gates.~\cite{Burkard2002,Klinovaja2012} Instead, combined with the proposals for realizing universal quantum logic gates in Ref.~\onlinecite{LA-Wu}, the controllability of the anisotropic exchange interaction which we find offers an {\it active} way to implement quantum computing with a strong-SOC system.

 \section{\label{II}The effective Hamiltonian of the nanowire DQD}

\begin{figure}
\includegraphics[width=0.4\textwidth]{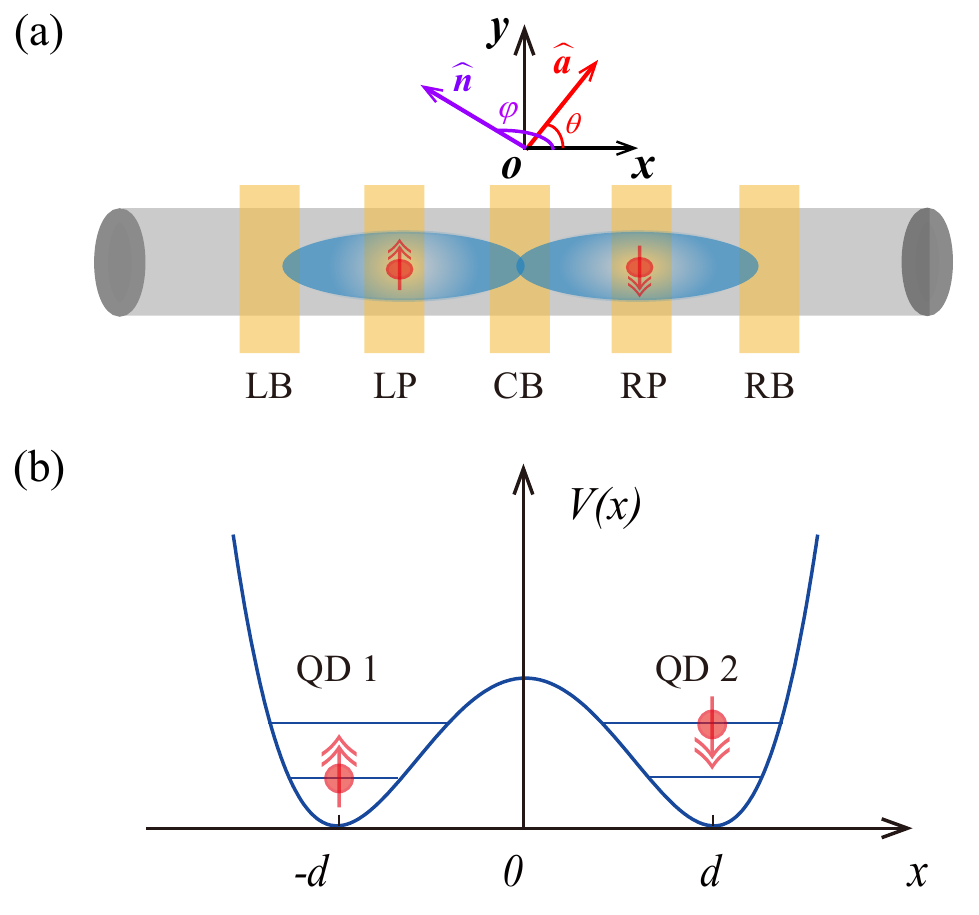}
\caption{(color online) (a) Schematic diagram of the nanowire double quantum dot (DQD) with a strong spin-orbit coupling (SOC). The gates  LB, CB, and RB define the barriers that form the DQD, and the plunger gates LP and RP control the electron occupancy of the individual QDs. The SOC vector $\hat{\mathbf{a}}=(\cos\theta,\sin\theta,0)$ and the magnetic-field direction $\hat{\mathbf{n}}=(\cos\varphi,\sin\varphi,0)$ are shown in the $x-y$ plane.   (b) The confinement potential along the wire direction (the $x$-axis). Each dot contains only a single electron, and because of the SOC the localized eigenstates of the QDs are quasi-spin states.}
\label{Fig1}
\end{figure}

 As shown in Fig.~\ref{Fig1}(a), we consider a semiconductor nanowire DQD with a strong SOC. For simplicity, the two QDs defined by the local gate electrodes are identical. The electron occupancy of each QD can be adjusted by changing the voltages on the electrodes.~\cite{Fasth2005,Wang2016}
Let the $x$-axis be the direction along the nanowire.  The axial confinement of the DQD can then be modeled as a double-well potential. As we show below, the details of this double-well potential are not important, since the coefficients of the effective Hamiltonian can actually be determined by fitting the experiments (see Sec.~\ref{III}). In order to obtain the explicit analytical form of the effective exchange interaction,  we approximate the potential as a quartic function
$V(x)=V_{0}[(x/d)^{2}-1]^{2}$,
with $2d$ being the interdot distance [see Fig.~\ref{Fig1}(b)].

In the presence of an external magnetic field applied in the $x-y$ plane, $\mathbf{B}=B(\cos\varphi,\sin\varphi,0)\equiv B \hat{\mathbf{n}}$, the Hamiltonian of an electron confined in the DQD reads~\cite{Liu2018,Flindt2006}
\begin{equation}
H_{0}(x)=\frac{p^{2}}{2m_{e}}+V(x)+\alpha_{\rm R}\sigma^{y}p+\alpha_{\rm D}\sigma^{x}p+\frac{g\mu_{B}B}{2}\sigma^{n},
\label{O-H}
\end{equation}
where $m_{e}$ is the effective electron mass, $p=-i\hbar\partial/\partial x$, $\alpha_{\rm R}$ ($\alpha_{\rm D}$) is the Rashba (Dresselhaus) SOC strength, $g$ is the  effective Land\'{e} factor, $\mu_{B}$ is the Bohr magneton, and $\sigma^{n}\equiv\hat{\boldsymbol{\sigma}}\cdot\hat{\mathbf{n}}$,  with the Pauli matrices $\hat{\boldsymbol{\sigma}}=(\sigma_{x},\sigma_{y},\sigma_{z})$.
Conveniently, by defining a new Pauli matrix $\sigma^{\rm a}\equiv\hat{\boldsymbol{\sigma}}\cdot\hat{\mathbf{a}}$, where the SOC vector is  $\hat{\mathbf{a}}=(\cos\theta,\sin\theta,0)$ with the angle
$\theta= {\rm arccot}(\alpha_{\rm D}/\alpha_{\rm R})$,
the SOC terms can be rewritten in a compact form, i.e., $\alpha_{\rm R}\sigma^{y}p+\alpha_{\rm D}\sigma^{x}p=\alpha\sigma^{\rm a}p$, where $\alpha=\sqrt{\alpha^{2}_{\rm D}+\alpha^{2}_{\rm R}}$.

Usually, the electronic eigenstates of a single QD in the coexistence of the Zeeman and  SOC terms are analytically obtained by perturbative approaches.~\cite{You2013,Trif2008}
In the context of strong SOC, it is optimal to perform an exact analysis of the SOC terms and treat the Zeeman term as a perturbation as long as the Zeeman splitting is much smaller than the orbital splitting, i.e., $g\mu_{B}B \ll \hbar\omega$,
where $\omega=\sqrt{8V_{0}/(d^{2}m_{e})}$. Meanwhile, in the case of small Zeeman splitting, only the lowest approximate Zeeman sublevels
in each dot are kept to facilitate the study of the low-energy dynamics of the electron, i.e., the Hund-Mulliken approximation. Let $|\Phi^{+}_{j}\rangle$ and $|\Phi^{-}_{j}\rangle$ ($j=1,~2$) denote the two lowest Zeeman sublevels of each QD.
In general, the localized eigenstates of the different dots are not orthogonal due to the nonzero overlaps among them.
Nevertheless, based on these four localized eigenstates, orthonormal basis states $|\Phi_{j\Uparrow}\rangle$ and $|\Phi_{j\Downarrow}\rangle$ ($j=1,~2$) can be constructed via the Schmidt orthogonalization.~\cite{SM}
Expanding the electron field operator in terms of the orthonormal basis states, $\Psi_{e}(x)=\sum_{j=1,~2;\sigma=\Uparrow,~\Downarrow}c_{j\sigma}|\Phi_{j\sigma}\rangle$,
we can write the second-quantization form of the Hamiltonian $H_{0}(x)$ in Eq.~(\ref{O-H})
as $H_{0}=\int \!dx \Psi^{\dagger}_{e}(x) H_{0}(x) \Psi^{}_{e}(x)=\sum_{j=1,2}\sum_{\sigma}\varepsilon^{}_{j\sigma}c^{\dagger}_{j\sigma}c^{}_{j\sigma}
\!+\!\sum_{\sigma}(t^{}_{\sigma}c^{\dagger}_{1\sigma}c^{}_{2\sigma}\!+ t'_{\sigma}c^{\dagger}_{1\sigma}c^{}_{2\bar{\sigma}}\!+
{\rm h.c.})$,
where $c^{\dagger}$ ($c$) is the electron creation (annihilation)
operator, $\varepsilon^{}_{j\sigma}$ represents the single-electron energy in each QD, while $t^{}_{\sigma}$ and $t'_{\sigma}$ represent, respectively, the spin-conserved and spin-flipped tunnelings between the two QDs (see Ref.~\onlinecite{SM} for their explicit expressions).

When there are two electrons confined in the nanowire DQD, keeping only the leading Coulomb-interaction terms, we can reduce the second-quantized Coulomb Hamiltonian $H_{\rm c}$ to
$H_{\rm c}=\frac{U'}{2}\sum_{j\neq j'}\sum_{\sigma\sigma'}n^{}_{j\sigma}n^{}_{j'\sigma'}
+\frac{U^{}}{2}\sum_{j}\sum_{\sigma}n^{}_{j\sigma}n^{}_{j\bar{\sigma}}$,
where $n^{}_{j\sigma}=c^{\dagger}_{j\sigma}c^{}_{j\sigma}$ is the particle number operator, and $U$ ($U'$) denotes the intradot (interdot) Coulomb repulsion. The total Hamiltonian of the system is $H=H_{0}+H_{\rm c}$.

Here we focus on the strong intradot Coulomb repulsion regime, i.e, $U-U'\gg |t^{}_{\sigma}|,|t'_{\sigma}|$, which means that each QD can only be occupied by one electron.
The effective Hamiltonian describing the two electrons can be simplified to~\cite{SM}
\begin{equation}
H_{\rm eff}=\Delta_{\rm z}(S^{z}_{ 1}+S^{z}_{ 2})+J\mathbf{S}^{}_{ 1}\cdot\mathbf{S}^{}_{ 2}+\mathbf{D}\cdot\mathbf{S}^{}_{ 1}\times\mathbf{S}^{}_{ 2}
+\mathbf{S}^{}_{ 1}\overleftrightarrow{\Gamma}\mathbf{S}^{}_{2},
\label{D-M}
\end{equation}
where $\mathbf{S}^{}_{j}=(1/2)\sum_{\sigma,\sigma'=
\Uparrow,\Downarrow}c^{\dagger}_{j\sigma}\hat{\boldsymbol{\sigma}}^{}_{\sigma\sigma'}c^{}_{ j \sigma'}$, $j=1,~2$, are the pseudo-spin operators defined by the orthonormal basis, and $\Delta^{}_{\rm z}=g\mu_{B}B f$ is the SOC-modified Zeeman splitting,~\cite{You2013} with
\begin{align}
f\equiv\sqrt{\cos^{2}(\varphi-\theta)+e^{-2x^{2}_{0}/x^{2}_{\rm so}}\sin^{2}(\varphi-\theta)},
\label{f-factor}
\end{align}
where $x_{0}\equiv\sqrt{\hbar/(m_{e}\omega)}$
is the ``Bohr" radius of the QDs, and $x_{\rm so}\equiv\hbar/(m_{e}\alpha)$ the spin-orbit length. The SOC-dependent exchange coupling strengths in Eq.~(\ref{D-M}) are
\begin{eqnarray}
J&=&J_{0}\cos^{2}\left(2d/x_{\rm so}\right),~~~
\mathbf{D}=J_{0}\sin\left(4d/x_{\rm so}\right)\hat{v},  \nonumber \\
\overleftrightarrow{\Gamma}\!&=&J_{0}\sin^{2}\left(2d/x_{\rm so}\right)(2\hat{v}\hat{v}-1),
\label{parameters}
\end{eqnarray}
with the bare exchange coupling strength $J_{0}=4(|t^{}_{\Uparrow}|^{2}+|t'_{\Uparrow}|^{2})/(U-U')$. The corresponding DM unit vector is
\begin{align}
 \hat{v}=\mathbf{e}_{z}\cos\gamma-\mathbf{e}_{x}\sin\gamma,
\label{DM vector}
\end{align}
where $\mathbf{e}_{\xi}$, $\xi=x,~y,~z$, represent the unit vectors in the three-dimensional space of pseudo-spins
and $\gamma$ is given by
\begin{align}
\gamma=\arccos\left[\cos(\varphi-\theta)/f\right],
\label{gamma}
\end{align}
with $\varphi-\theta$ being the angle between the SOC direction $\mathbf{a}$ and the applied magnetic field, and with
$f$ given in Eq.~(\ref{f-factor}). 
Interestingly, the two-electron exchange coupling in Eq.~(\ref{D-M}) is shown to possess Moriya's anisotropic superexchange interaction.~\cite{Aharony1992,Shekhtman1993} Also, it should be noted that the specific form of the confinement potential between the two QDs does not alter the exchange-interaction structure in Eq.~(\ref{D-M}), but it affects the magnitude of $J_{0}$.
In practice, for a realistic nanowire DQD, the value of $J_{0}$ can be detected experimentally (see below).

\section{\label{III}The role of the Zeeman term}

In the absence of the external magnetic field, the Hamiltonian describing an electron in the DQD reads
$H'_{0}(x)=p^{2}/(2m_{e})+V(x)+\alpha_{\rm R}\sigma^{y}p+\alpha_{\rm D}\sigma^{x}p$.
This Hamiltonian possesses time-reversal symmetry, i.e., $(i\sigma^{y}K)H'_{0}(i\sigma^{y}K)^{-1}=H'_{0}$, where $K$ is the complex conjugate operator,
so that there are degenerate states in this case (Kramers pairs). In the presence of nonzero magnetic field, we can also obtain the effective Hamiltonian of the two electrons in the DQD. However, because of the Kramers degeneracy in the present case, the direction of the anisotropic exchange interaction is not determined, and can be chosen arbitrarily. Here we assume that the DM vector $\hat{v}$ takes the same form as Eq.~(\ref{DM vector}) to facilitate the study, and the effective Hamiltonian of the two electrons is
$H'_{\rm eff}=J\mathbf{S}^{}_{1}\cdot\mathbf{S}^{}_{2}+\mathbf{D}\cdot\mathbf{S}^{}_{1}\times\mathbf{S}^{}_{2}
+\mathbf{S}^{}_{1}\overleftrightarrow{\Gamma}\mathbf{S}^{}_{2}$,
with the parameters given in Eq.~(\ref{parameters}).

Via a unitary transformation, the above anisotropic exchange Hamiltonian $H'_{\rm eff}$ can be mapped onto an isotropic Heisenberg Hamiltonian,~\cite{Aharony1992,Kaplan1983}
$\widetilde{H}'_{\rm eff}= J_{0} \widetilde{\mathbf{S}}^{}_{1}\cdot\widetilde{\mathbf{S}}^{}_{2}$,
where the spin operators $\widetilde{\mathbf{S}}_{1}$ and $\widetilde{\mathbf{S}}_{2}$ are obtained by rotating $\mathbf{S}_{1}$ and $\mathbf{S}_{2}$ around the DM vector $\hat{v}$  with angles $-\vartheta$ and $\vartheta$, respectively,
\begin{align}
\widetilde{\mathbf{S}}^{}_{1}\equiv&\exp(i\vartheta\hat{v}\cdot\hat{\mathbf{S}})\mathbf{S}^{}_{ 1} \exp(-i \vartheta\hat{v}\cdot\hat{\mathbf{S}}), \nonumber\\
\widetilde{\mathbf{S}}^{}_{2}\equiv&\exp(-i\vartheta\hat{v}\cdot\hat{\mathbf{S}})\mathbf{S}^{}_{ 2}\exp(i \vartheta\hat{v}\cdot\hat{\mathbf{S}}),
 \end{align}
where $\vartheta=2d/x_{\rm so}$ and $\hat{\mathbf{S}}=(1/2)\hat{\boldsymbol{\sigma}}$.

When an external Zeeman field is applied, time-reversal symmetry is broken and the two-electron Hamiltonian takes the form of Eq.~(\ref{D-M}).
If we still perform the rotation for Eq.~(\ref{D-M}) as done above, the rotated Hamiltonian becomes
$\widetilde{H}_{\rm eff}=\widetilde{H}_{\rm 0}+\Delta\widetilde{H}$, with
\begin{eqnarray}
\widetilde{H}_{0}&=&J_{0}\widetilde{\mathbf{S}}_{1}\cdot\widetilde{\mathbf{S}}_{2}
+\Delta_{\rm z}\frac{\mathbf{B}_{1}+\mathbf{B}_{2}}{2B}\cdot(\widetilde{\mathbf{S}}_{1}
+\widetilde{\mathbf{S}}_{2}), \nonumber \\
\Delta\widetilde{H}&=&\Delta_{\rm z}\frac{\mathbf{B}_{1}-\mathbf{B}_{2}}{2B} \cdot(\widetilde{\mathbf{S}}_{1}
-\widetilde{\mathbf{S}}_{2}).
\label{e-ms}
\end{eqnarray}
The uniform external magnetic field now becomes an effective {\it inhomogeneous} magnetic field, with the local effective magnetic fields in the two QDs given by $\mathbf{B}_{1}=B(\beta_{x},\beta_{y},\beta_{z})$ and $\mathbf{B}_{2}=B(\beta_{x},-\beta_{y},\beta_{z})$, where $\beta_{x}=(\cos\vartheta-1)\sin\gamma\cos\gamma$, $\beta_{y}=\sin\vartheta\sin\gamma$, and
$\beta_{z}=\cos^{2}\gamma+\sin^{2}\gamma\cos\vartheta$.
It is easy to find that the eigenstates of $\widetilde{H}_{0}$ are the
singlet and triplet states $|\rm{S}_{0}\rangle=(1/\sqrt{2})(|\uparrow\downarrow\rangle-|\downarrow\uparrow\rangle)$, $|\rm{T}_{0}\rangle=(1/\sqrt{2})(|\uparrow\downarrow\rangle+|\downarrow\uparrow\rangle)$, $|\rm{T}_{-}\rangle=|\downarrow\downarrow\rangle$, and  $|\rm{T}_{+}\rangle=|\uparrow\uparrow\rangle$, with the spin direction determined by the average field direction $\hat{\mathbf{w}}\equiv(\mathbf{B}_{1}+\mathbf{B}_{2})/(2B)$:
$\hat{\mathbf{w}}\cdot\hat{\boldsymbol{\sigma}}|\rm{T}_{\pm}\rangle=\pm\sqrt{1-\beta^{2}_{y}}|\rm{T}_{\pm}\rangle$.
However, $\Delta\widetilde{H}$ can lead to the mixing of the triplet and singlet states. Expanding the Hamiltonian $\widetilde{H}_{\rm eff}$ in the subspace spanned by $|\rm{T}_{+}\rangle$, $|\rm{T}_{0}\rangle$, $|\rm{T}_{-}\rangle$ and $|\rm{S}_{0}\rangle$, we obtain
\begin{align}
\widetilde{H}_{\rm eff}=&\begin{pmatrix}
\frac{J_{0}}{4}+\Delta'_{\rm z} &0 & 0 &-i\frac{\sqrt{2}}{2}\beta_{y}\Delta_{\rm z} \\
0 & \frac{J_{0}}{4} &0  &0 \\
0 & 0 & \frac{J_{0}}{4}-\Delta'_{\rm z} & -i\frac{\sqrt{2}}{2}\beta_{y}\Delta_{\rm z}\\
i\frac{\sqrt{2}}{2}\beta_{y}\Delta_{\rm z}&0& i\frac{\sqrt{2}}{2}\beta_{y}\Delta_{\rm z}&-\frac{3J_{0}}{4}
\end{pmatrix},
\label{M-P}
\end{align}
with $\Delta'_{\rm z}=\sqrt{1-\beta^{2}_{y}}\Delta_{\rm z}$.

\begin{figure}
\includegraphics[width=0.40\textwidth]{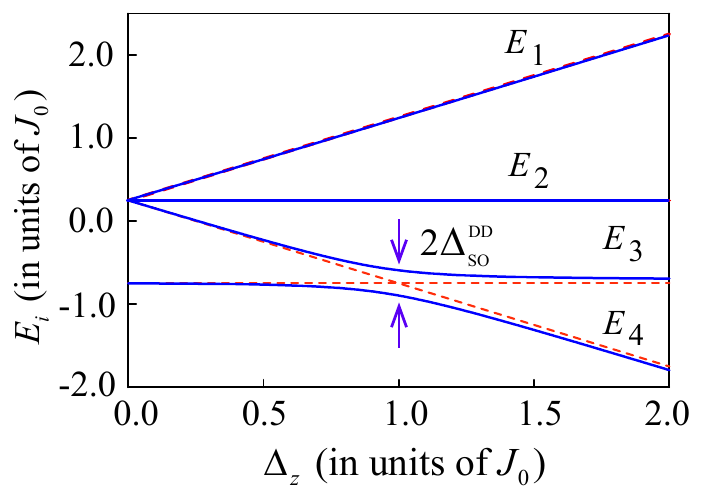}
\caption{(color online) The energy spectrum of the nanowire DQD versus the Zeeman-field splitting $\Delta_{\rm z}$ for different values of the inhomogeneity $\Delta B$. The dashed (red) curves represent the case when $\Delta B=0$. The solid (blue) curves are for $\Delta B=0.21$,  where $2\Delta^{\rm DD}_{\rm SO}$ denotes the energy gap between the levels $|\Phi_{3}\rangle$ and $|\Phi_{4}\rangle$ at the anticrossing point when $\Delta_{\rm z}\simeq J_{0}$.   }
\label{Fig2}
\end{figure}

Below we show that the effect of the SOC can be reflected by the inhomogeneity of the effective magnetic field,
 \begin{align}
\Delta B\equiv
\frac{|\mathbf{B}_{1}-\mathbf{B}_{2}|}{2B}=|\beta_{y}|.
\label{B-D}
\end{align}
When there is no difference between the two local effective fields, i.e., $\Delta B=0$, obviously the eigenstates of the Hamiltonian $\widetilde{H}_{\rm eff}$  are the singlet and triplet states $|\Phi_{1}\rangle=|\rm{T}_{+}\rangle$,  $|\Phi_{2}\rangle=|\rm{T}_{0}\rangle$, $|\Phi_{3}\rangle=|\rm{T}_{-}\rangle$, and $|\Phi_{4}\rangle=|\rm{S}_{0}\rangle$. Note, though, that there is an energy-level crossing between the singlet and triplet states, i.e., $|\Phi_{3}\rangle$ and $|\Phi_{4}\rangle$, at a critical magnetic field $B_{0}$, where the Zeeman splitting is
$\Delta_{\rm z}=J_{0}$
(see Fig.~\ref{Fig2}). However, in the case of two different local effective fields, i.e., $\Delta B \neq 0$, the energy-level crossing is avoided around the critical magnetic field and there is an anticrossing between these two levels 
(see Fig.~\ref{Fig2}).  For a nanowire DQD with the spin-orbit length $x_{\rm so}\gg d,x_{0}$, the singlet-triplet splitting at the anticrossing point can be analytically written as
$2\Delta^{\rm DD}_{\rm SO}= \sqrt{2}J_{0}\Delta B$.

To fit the model with the real system, the parameters of the nanowire DQD used in the following calculations are taken from Ref.~\onlinecite{Perge2012}, with $2d=50$~nm and  $x_{0}\simeq 30$~nm.
Moreover, it was found that $g\simeq 33$~\cite{note1} and $B_{0}\simeq 13.3$~mT for the InSb nanowire DQD. From these experimentally-detected values, basing on $\Delta_{\rm z}=J_{0}$, we can deduce the bare exchange coupling strength $J_{0}\simeq 24.6~\rm{\mu eV}$. The specific value of the energy splitting $\Delta^{\rm DD}_{\rm SO}$ depends on the magnetic-field direction and the SOC in the nanowire, as explained below.

\section{\label{IV}Dependence on the magnetic-field direction}
It is known that the magnetic-field direction plays an important role in observing the SOC effects in QDs.~\cite{Nowak2011,Falko2005,Scarlino2014,Stepanenko2012}
Below we study the influence of the magnetic-field direction on the two-electron exchange interaction in the nanowire DQD.

\begin{figure}
\centering
\includegraphics[width=0.45\textwidth]{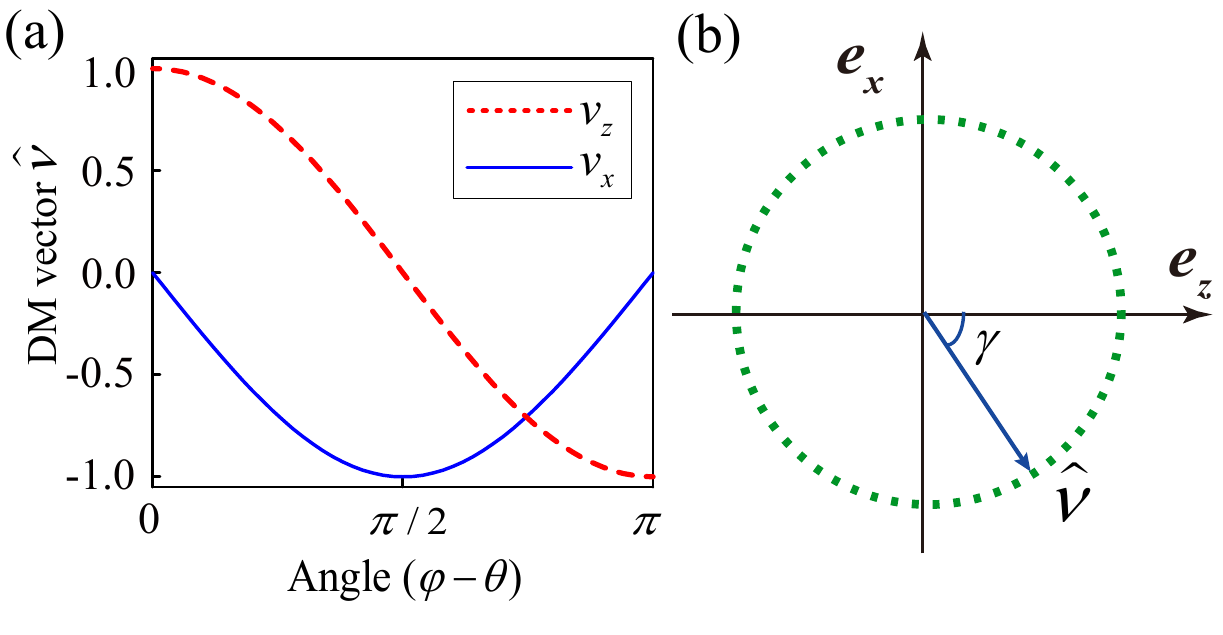}
\caption{(color online) (a) Components of the DM vector $\hat{v}$ as a function of the angle $\varphi-\theta$, for $x_{0}=30$~nm and $x_{\rm so}=200$~nm. (b) Schematic diagram of the DM vector $\hat{v}$ in the $\mathbf{e}_{x}-\mathbf{e}_{z}$ plane, with the azimuthal angle $\gamma=\arccos\left[\cos(\varphi-\theta)/f\right]$.}
\label{Fig3}
\end{figure}

From the analytical expressions for the DM vector $\hat{v}$ in Eq.~(\ref{DM vector}) and the angle $\gamma$ in Eq.~(\ref{gamma}), it is easy to find that the exchange anisotropy direction in the DQD can be manipulated by regulating the direction of the magnetic field.
In Fig.~\ref{Fig3}(a), the components of the DM vector $\hat{v}$ versus the angle $\varphi-\theta$ are plotted.
For example, when the magnetic-field direction is perpendicular to the SOC direction, i.e., $\varphi-\theta=90^{\circ}$ or $270^{\circ}$,
the DM interaction points along the $-\mathbf{e}_{x}$ or $\mathbf{e}_{x}$  direction.
If the two directions are parallel, i.e., $\varphi-\theta=0^{\circ}$ or $180^{\circ}$, 
the DM interaction is along the $\mathbf{e}_{z}$ or $-\mathbf{e}_{z}$ direction.
However, in this case, the SOC seems to have a trivial contribution to the anisotropic exchange interaction because the effective magnetic field is homogeneous ($\Delta B=0$).
In short, we can continuously rotate the DM vector in the $\mathbf{e}_{x}-\mathbf{e}_{z}$ plane by just varying the magnetic-field direction, as shown in Fig.~\ref{Fig3}(b).

\begin{figure}
\centering
\includegraphics[width=0.43\textwidth]{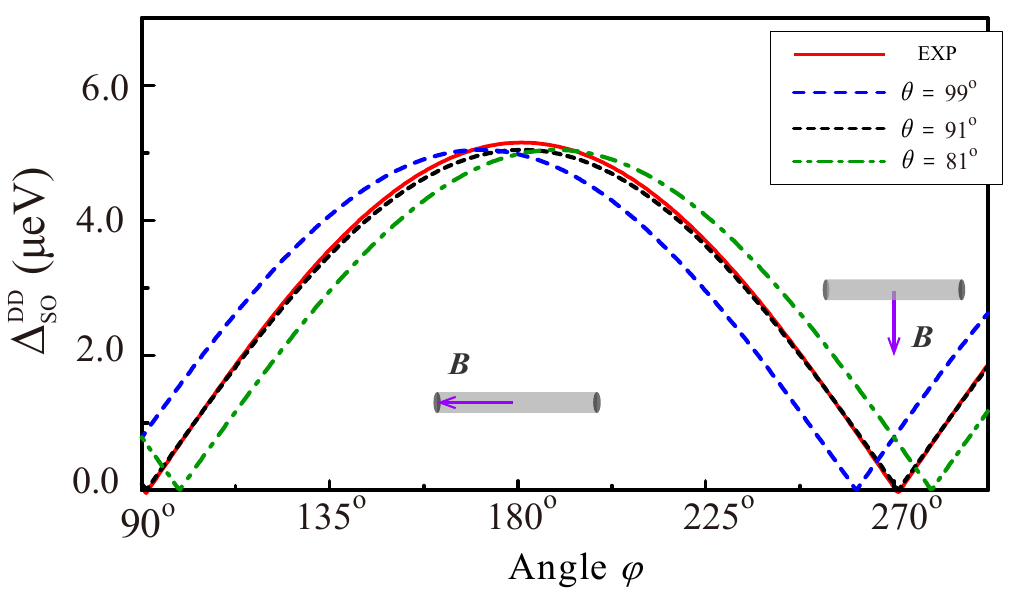}
\caption{(color online) The magnitude of the energy gap $\Delta^{\rm DD}_{\rm SO}$ as a function of the magnetic-field direction angle $\varphi$.
The solid (red) line corresponds to the experimental data in Ref.~\onlinecite{Perge2012}.
The other three lines represent the theoretical results based on Eq.~(\ref{DD-SO-1}) under different values of $\theta$, with $J_{0}\simeq 24.6~\rm{\mu eV}$, $2d=50$~nm, and  $x_{\rm so}=165~{\rm nm}$.}
\label{Fig4}
\end{figure}

Furthermore, based on the rotated $\widetilde{H}_{\rm eff}$  and for the SOC-dependent factor $f\simeq 1$, we obtain a specific relationship between $\Delta^{\rm DD}_{\rm SO}$ and $\varphi$ at the anticrossing point,
 \begin{align}
 \Delta^{\rm DD}_{\rm SO}=\frac{\sqrt{2}}{2}J_{0}\sin(2d/x_{\rm so})|\sin(\varphi-\theta)|.
 \label{DD-SO-1}
 \end{align}
Interestingly, the cosine dependence of the energy gap $\Delta^{\rm DD}_{\rm SO}$
on the magnetic-field direction angle $\varphi$ has indeed
been detected experimentally, cf.~Fig.~4(i) in Ref.~\onlinecite{Perge2012}, i.e.,
\begin{align}
\Delta^{\rm DD}_{\rm SO}=\Delta_{\rm SO}|\cos(\varphi-\varphi_{0})|,
\label{DD-SO-2}
\end{align}
with the fitting parameters $\Delta_{\rm SO}\simeq 5.2~\mu$eV  and $\varphi_{0}\simeq 1^{\circ}$.
By comparing the experimental fitting function with the analytical expression of $\Delta^{\rm DD}_{\rm SO}$ in Eq.~(\ref{DD-SO-1}), we can obtain the formulae for the spin-orbit length $x_{\rm so}$ and the spin-orbit angle $\theta$,
\begin{align}
x_{\rm so}=\frac{2d}{\arcsin(\sqrt{2}\Delta_{\rm SO}/J_{0})}, ~~~~~~
\theta=\frac{\pi}{2}+\varphi_{0}.
\label{so-ex}
\end{align}
Using the specific values of the fitting parameters and the bare exchange coupling strength, we find the magnitudes of the spin-orbit parameters,
$x_{\rm so}=165~{\rm nm}$, and $\theta_{\rm ex}=91^{\circ}$.~\cite{note2} The agreement between the experimental observations and the theoretical results is shown in Fig.~\ref{Fig4}.

For the spin-orbit length in an InSb nanowire DQD, the discrepancy between the theoretical result $x_{\rm so}=165~{\rm nm}$ and the experimental estimate in Ref.~\onlinecite{Perge2012}, $l_{\rm so}=230$~nm, mainly originates from the different quantitative methods. In the experiment, the spin-orbit length was quantified using an approximation method,~\cite{supplement2012} $(2d/l_{\rm so})\approx (\Delta_{\rm SO}/J_{0})$,  and then from Eq.~(\ref{so-ex}) the ratio of $l_{\rm so}$ to $x_{\rm so}$ can be identified as $l_{\rm so}/x_{\rm so}\simeq \sqrt{2}$. Using the definition of the spin-orbit angle  $\theta\equiv{\rm arccot}(\alpha_{\rm D}/\alpha_{\rm R})$, in the absence of the Dresselhaus SOC, i.e., $\theta=90^{\circ}$ or $270^{\circ}$, it follows from Eq.~(\ref{DD-SO-1}) that the magnitude of $\Delta^{\rm DD}_{\rm SO}$ reaches its maximal (minimal) value when
the magnetic field is parallel (perpendicular) to the nanowire axis. Obviously, it is the presence of a small Dresselhaus SOC that gives rise to $\theta_{\rm ex}=91^{\circ}$. The absolutely dominant role of the Rashba SOC, which would imply $\theta_{\rm ex}=90^{\circ}$, was predicted by the symmetry analysis of the nanowire DQD in Ref.~\onlinecite{Perge2012}. Our fit, which does give a small Dresselhaus contribution, must result from deviations of the finite sample from the ideal crystal.~\cite{notes}
On the other hand, the consistency between our theoretical results (Fig.~\ref{Fig4}) and the experimental detection~\cite{Perge2012} validates the controllability of the exchange interaction in the nanowire DQD by varying the direction of the external magnetic field.

\section{\label{V}Conclusions}
We have studied the two-electron anisotropic exchange interaction in a nanowire DQD under the influence of a strong SOC and a Zeeman field.
As in the case of zero magnetic field, the exchange interaction can be mapped onto an isotropic Heisenberg interaction, but the uniform external magnetic field becomes then an effective inhomogeneous field and the inhomogeneity of this effective magnetic field reflects the SOC strength.
Also, we reveal the controllability of the anisotropic exchange interaction by tuning the direction of the external magnetic field and obtain an analytical expression for the dependence of the singlet-triplet splitting on the magnetic-field direction, as detected in an InSb nanowire DQD.~\cite{Perge2012}
Our theory provides a tool to explore the novel properties of the exchange interaction in a nanowire DQD under the strong SOC and also offers a complete method to experimentally detect separately the Rashba and Dresselhaus SOCs in the nanowire.

\begin{acknowledgments}
We thank R. I. Shekhter, R. Li and S. Gurvitz for useful discussions. This work is supported by the National Key Research and Development Program of China (Grant No. 2016YFA0301200) and the NSFC (Grant No.~11774022). O. E. W. and A. A. are partially supported by the Israel Science Foundation (I. S. F.), by the infrastructure program of Israel Ministry of Science and Technology under contract 3-11173, and by the Pazy Foundation.
\end{acknowledgments}


\begin{thebibliography}{99}
\bibitem{notes}In the actual InSb nanowire, the $\sim$$1^\circ$ deviation of the spin-orbit angle $\theta_{\rm ex}$ from $90^\circ$ is most likely due to uncertainties present in the measurements. The Dresselhaus SOC  may also result from the finiteness of the sample, due to the influence of its boundaries or the misalignment of the lattice field.
\end{thebibliography}
\end{document}


\title{Supplemental Material for \\Control of the Two-Electron Exchange Interaction in a Nanowire Double Quantum Dot}

\author{Zhi-Hai Liu }
\affiliation{Interdisciplinary Center of Quantum Information and Department of Physics, Zhejiang University, Hangzhou 310027, China}
\affiliation{Quantum Physics and Quantum Information Division, Beijing Computational Science Research Center, Beijing 100193, China}

\author{O. Entin-Wohlman}
\affiliation{Raymond and Beverly Sackler School of Physics and Astronomy, Tel Aviv University, Tel Aviv 69978, Israel}
\affiliation{Physics Department, Ben Gurion University, Beer Sheva 84105, Israel}

\author{A. Aharony}
\affiliation{Raymond and Beverly Sackler School of Physics and Astronomy, Tel Aviv University, Tel Aviv 69978, Israel}
\affiliation{Physics Department, Ben Gurion University, Beer Sheva 84105, Israel}

\author{J. Q. You}
\affiliation{Interdisciplinary Center of Quantum Information and Department of Physics, Zhejiang University, Hangzhou 310027, China}
\affiliation{Quantum Physics and Quantum Information Division, Beijing Computational Science Research Center, Beijing 100193, China}

\maketitle

\section{Derivation of the orthonormal basis states}

First, we calculate the localized eigenstates that construct the orthonormal
basis states.
Near the minima of the axial confinement potential $V(x)=V_{0}[(x/d)^{2}-1]^{2}$, the confining potential can be expanded harmonically as $V_{1/2}(x)=(m_{e}\omega^{2}/2)(x\pm d)^{2}$,
where $\omega=\sqrt{8V_{0}/(d^{2}m_{e})}$.
Then, the Hamiltonian in Eq.~(1) of the main text becomes
\begin{align}
H_{1/2}(x)=&\frac{p^{2}}{2m_{e}}+\frac{m_{e}\omega^{2}}{2}\left(x\pm d\right)^{2}+\alpha_{\rm R}\sigma^{y}p+\alpha_{\rm D}\sigma^{x}p\nonumber\\
&~~~~~~ +\frac{g\mu_{B}B}{2}\sigma^{n}.
\label{L-H}
 \end{align}
Based on the local harmonic potential, in the absence of the external magnetic field, the localized electron wavefunctions in a single nanowire dot can be solved analytically. Let $|\Phi_{jn\sigma}\rangle$ denote the eigenstates of the $j$th QD without the external magnetic field, in which  $n=0,1,2,...$ represent the orbital quantum numbers and $\sigma=\uparrow,\downarrow$ represent the electron spin states. The explicit forms of the two degenerate states $|\Phi_{jn\uparrow}\rangle$ and $|\Phi_{jn\downarrow}\rangle$ are given by
\begin{align}
|\Phi_{1/2n\uparrow}\rangle=&e^{-i(x\pm d)/x_{\rm so}}\psi^{}_{n}\left(x\pm d\right)|\uparrow_{a}\rangle,\nonumber \\
|\Phi_{1/2n\downarrow}\rangle=&e^{i(x\pm d)/x_{\rm so}}\psi^{}_{n}\left(x\pm d\right)|\downarrow_{a}\rangle,
\end{align}
where $x_{\rm so}=\hbar/(m_{e}\alpha)$ is the spin-orbit length, $\psi^{}_{n}(x)$ is the eigenstate of a one-dimensional harmonic oscillator with eigenvalue $\varepsilon_{n}=(n+\frac{1}{2})\hbar\omega$,
and $|\uparrow_{\rm a}\rangle$ and $|\downarrow_{\rm a}\rangle$ are the two eigenstates of $\sigma^{\rm a}$: $\sigma^{\rm a}|\uparrow_{\rm a }\rangle=|\uparrow_{\rm a }\rangle$ and $\sigma^{\rm a}|\downarrow_{\rm a }\rangle=-|\downarrow_{\rm a }\rangle$.

The degeneracy of the localized states is lifted in the presence of the external magnetic field. The Zeeman term can be treated as a perturbation in the calculations as long as $\xi\equiv \frac{g\mu_{\rm B}B}{\hbar\omega} \ll 1 $.~\cite{Nowak2013,You2013} In first-order perturbation theory, the Zeeman term mixes the orbital states $|\Phi_{j0\sigma}\rangle$ and $|\Phi_{jn\sigma'}\rangle$ ($n\neq 0$), with the degree of the hybridization being proportional to $\frac{\xi}{n}(\frac{x_{0}}{x_{\rm so}})^{n}$, where $x_{0}=\sqrt{\frac{\hbar}{m_{e}\omega}}$ is the ``Bohr" radius of the QDs. To facilitate the study,
it is reasonable to consider only the mixing of the two lowest orbital states in the context of a small magnetic field and $x_{0}< x_{\rm so}$. Thus, the lowest Zeeman sublevels of the localized QDs can be approximated by
\begin{widetext}
\begin{align}
|\Phi^{+}_{j}\rangle=&\frac{\cos(\gamma/2)}{\sqrt{\chi^{2}+1}}
\left(|\Phi^{}_{j0\uparrow}\rangle+\chi|\Phi^{}_{j1\downarrow}\rangle\right)
-i\frac{\sin(\gamma/2)}{\sqrt{\chi^{2}+1}}
\left(|\Phi^{}_{j0\downarrow}\rangle+\chi|\Phi^{}_{j1\uparrow}\rangle\right), \nonumber\\
|\Phi^{-}_{j}\rangle=&-\frac{\sin(\gamma/2)}{\sqrt{\chi^{2}+1}}
\left(|\Phi^{}_{j0\uparrow}\rangle+\chi|\Phi^{}_{j1\downarrow}\rangle\right)
-i\frac{\cos(\gamma/2)}{\sqrt{\chi^{2}+1}}
\left(|\Phi^{}_{j0\downarrow}\rangle+\chi|\Phi^{}_{j1\uparrow}\rangle\right),
\label{loc}
\end{align}
\end{widetext}
with the angle $\gamma$ given in Eq.~(6) of the main text and the degree of the orbital hybridization being $\chi=\frac{\sqrt{2}}{2}\frac{x_{0}}{x_{\rm so}}\xi e^{-x_{0}^{2}/x^{2}_{\rm so}}\sin(\varphi-\theta)$.
The energy difference between the quasi-spin states $|\Phi^{+}_{ 1/2}\rangle$ and
$|\Phi^{-}_{1/2}\rangle$ is modified by the spin-orbit coupling (SOC) effect,
\begin{equation}
\Delta^{}_{\rm z}=g\mu_{\rm B}Bf,
\end{equation}
with $f=\sqrt{\cos^{2}(\varphi-\theta)+e^{-2x^{2}_{0}/x^{2}_{\rm so}}\sin^{2}(\varphi-\theta)}$ being a SOC-dependent factor.~\cite{You2013}
Therefore, we have obtained four localized wavefunctions (two for the left dot and two for the right dot).
These four localized states $|\Phi^{\pm}_{ 1}\rangle$ and $|\Phi^{\pm}_{2}\rangle$ are not orthogonal to each other, because
\begin{align}
\langle\Phi^{-}_{ 1}|\Phi^{-}_{2}\rangle=&\langle\Phi^{+}_{1}|\Phi^{+}_{2}\rangle^{\ast}=s_{d},  \nonumber\\
\langle\Phi^{-}_{ 1}|\Phi^{+}_{ 2}\rangle=&\langle\Phi^{+}_{ 1}|\Phi^{-}_{2}\rangle=s_{x},
\end{align}
with
\begin{widetext}
\begin{align}
s_{ d }=&\frac{e^{-d^{2}/x^{2}_{0}}}{\chi^{2}+1}\left[\cos\left(\frac{2d}{x_{\rm so}}\right)-i\cos\gamma\sin\left(\frac{2d}{x_{\rm so}}\right)\right]+\chi^{2}\frac{e^{-d^{2}/x^{2}_{0}}}{\chi^{2}+1}\frac{x^{2}_{0}-2d^{2}}{x^{2}_{0}}\left[\cos\left(\frac{2d}{x_{\rm so}}\right)
+i\cos\gamma\sin\left(\frac{2d}{x_{\rm so}}\right)\right] \nonumber \\
&+2\sqrt{2}\chi\frac{d}{x_{0}}\frac{e^{-d^{2}/x^{2}_{0}}}{\chi^{2}+1}\sin\gamma\sin\left(\frac{2d}{x_{\rm so}}\right), \nonumber\\
s_{ x}=&  -i\frac{e^{-d^{2}/x^{2}_{0}}}{\chi^{2}+1}\left(1-\chi^{2}\frac{x^{2}_{0}-2d^{2}}{x^{2}_{0}}\right)\sin\gamma\sin\left(\frac{2d}{x_{\rm so}}\right) -2\sqrt{2}\chi\frac{e^{-d^{2}/x^{2}_{0}}}{\chi^{2}+1} \frac{d}{x_{0}}\cos\gamma\sin\left(\frac{2d}{x_{0}}\right),
\label{overlaps}
\end{align}
\end{widetext}
which are usually not equal to zero.

According to the Schmidt diagonalization, we define two orthonormal basis states
\begin{align}
|\Phi^{}_{ 1 \Downarrow}\rangle=&\frac{1}{\sqrt{\tau_{1}}}\left(|\Phi^{-}_{ 1}\rangle-\varsigma|\Phi^{-}_{2}\rangle\right),  \nonumber \\
|\Phi^{}_{2 \Downarrow}\rangle=&\frac{1}{\sqrt{\tau_{1}}}\left(|\Phi^{-}_{2 }\rangle-\varsigma^{\ast}|\Phi^{-}_{ 1 }\rangle\right),
\label{down}
\end{align}
where
\begin{align}
&\varsigma=\frac{1}{s_{d}}\left(1-\sqrt{1-|s_{ d}|^{2}}\right),  \nonumber\\
&\tau_{1}=1+|\varsigma|^{2}-2{\rm Re}\left[\varsigma s_{ d}\right],
\end{align}
with the explicit expressions of $s_{d}$ and $s_{x}$ given in Eq.~(\ref{overlaps}).
In addition, we introduce two auxiliary states which are orthogonal to
$|\Phi_{1 \Downarrow}\rangle$ and $|\Phi_{ 2 \Downarrow}\rangle$,
\begin{align}
|\Phi^{a}_{1 \Uparrow}\rangle=&\frac{1}{\sqrt{\tau^{a}_{2}}}\left(|\Phi^{+}_{1 }\rangle-\xi^{a}_{1}|\Phi^{}_{ 1\Downarrow }\rangle-\xi^{a}_{2}|\Phi_{ 2\Downarrow }\rangle\right), \nonumber \\
|\Phi^{a}_{2 \Uparrow}\rangle=&\frac{1}{\sqrt{\tau^{a}_{2}}}\left(|\Phi^{+}_{2 }\rangle-\xi^{a\ast}_{2}|\Phi^{}_{ 1\Downarrow }\rangle-\xi^{a\ast}_{1}|\Phi_{ 2\Downarrow }\rangle\right),
\end{align}
with $\xi^{a}_{1}=-s^{\ast}_{x}\varsigma^{\ast}/\sqrt{\tau_{1}}$, $\xi^{a}_{2}=s^{\ast}_{x}/\sqrt{\tau_{1}}$, and $\tau^{a}_{2}=1-|\xi^{a}_{1}|^{2}-|\xi^{a}_{2}|^{2}$. Note that there exists an overlap between $|\Phi^{a}_{1 \Uparrow}\rangle$ and $|\Phi^{a}_{2 \Uparrow}\rangle$. Nevertheless, we can derive the other two orthonormal basis states using these auxiliary states, just in the same way as for $|\Phi_{ 1 \Downarrow}\rangle$ and $|\Phi_{2 \Downarrow}\rangle$.  In terms of the localized states, the other two orthonormal basis states can be written as
\begin{align}
|\Phi^{}_{ 1 \Uparrow}\rangle=&\frac{1}{\sqrt{\tau_{2}}}\left(|\Phi^{+}_{ 1 }\rangle-\xi_{1}|\Phi^{+}_{ 2}\rangle+\xi_{2}|\Phi^{-}_{ 1}\rangle-\xi_{3}|\Phi^{-}_{ 2 }\rangle\right), \nonumber \\
|\Phi^{}_{2 \Uparrow}\rangle=&\frac{1}{\sqrt{\tau_{2}}}\left(|\Phi^{+}_{2 }\rangle-\xi^{\ast}_{1}|\Phi^{+}_{ 1 }\rangle+\xi^{\ast}_{2}|\Phi^{-}_{ 2}\rangle-\xi^{\ast}_{3}|\Phi^{-}_{ 1 }\rangle\right),
\label{up}
\end{align}
where
\begin{align}
  \xi_{1}=&\frac{1}{\varrho}\left(1-\sqrt{1-|\varrho|^{2}}\right), \nonumber\\
  \xi_{2}=&\frac{1}{\tau^{}_{1}}\left[-2\varsigma^{\ast} s^{}_{ x}+\xi^{}_{1} s^{}_{ x}(1+|\varsigma|^{2})\right], \nonumber \\
  \xi_{3}=&~~\frac{1}{\tau^{}_{1}}\left[2\xi^{}_{1}\varsigma s^{}_{ x}-s^{}_{x}(1+|\varsigma|^{2})\right], \\
  \tau_{2}=&\frac{\tau^{}_{1}}{\tau^{}_{1}-s^{2}_{ x}(|\varsigma|^{2}+1)} \left\{1+|\xi^{}_{1}|^{2}-2{\rm Re}[\xi^{}_{1}\varrho]\right\} \nonumber,
\end{align}
 with $\varrho=(s^{\ast}_{d}\tau_{1}+2\varsigma s^{2}_{x})/[\tau_{1}-s^{2}_{ x}(|\varsigma|^{2}+1)]$.\\

\section{ The second-quantization form of the single-electron Hamiltonian}

Based on the orthonormal basis states, we can obtain the expansion of the electron field operator  $\Psi_{e}(x)=\sum_{j=1,~2;\sigma=\Uparrow,~\Downarrow}c_{j\sigma}|\Phi_{j\sigma}\rangle$ and
derive the second-quantization form of the single-electron Hamiltonian $H_{0}(x)$ in Eq.~(1) of the main text,
\begin{align}
H_{0}= & \int dx \Psi^{\dagger}_{e}(x) H_{0}(x) \Psi^{}_{e}(x) \nonumber \\
=&H_{\rm e}+H_{\rm t},
\label{s-q}
\end{align}
with
\begin{align}
H_{\rm e}=&\sum_{j=1,2}\sum_{\sigma}\varepsilon^{}_{j\sigma}c^{\dagger}_{j\sigma}c^{}_{j\sigma}, \nonumber\\
H_{\rm t}=&\sum_{\sigma}\left(t^{}_{\sigma}c^{\dagger}_{1\sigma}c^{}_{2\sigma}+t'_{\sigma}c^{\dagger}_{1\sigma}c^{}_{2\bar{\sigma}}+\rm h.c.\right),
\end{align}
where the parameters of the Hamiltonian can be calculated using the orthonormal basis states,
\begin{align}
t^{}_{\sigma}=&\langle\Phi^{}_{1\sigma}|H_{0}(x)|\Phi^{}_{2\sigma}\rangle,  \nonumber\\
t'_{\sigma}=&\langle\Phi^{}_{1\sigma}|H_{0}(x)|\Phi^{}_{2\bar{\sigma}}\rangle, \\
\varepsilon^{}_{j\sigma}=&\langle\Phi^{}_{j\sigma}|H_{0}(x)|\Phi^{}_{j\sigma}\rangle. \nonumber
\end{align}
To facilitate the calculations, we write the Hamiltonian $H_{0}(x)$ as $H_{0}(x)=H_{j}(x)+\Delta V_{j}(x)$, with $\Delta V_{j}\equiv V(x)-V_{j}(x)$ denoting the potential deviation from $V_{j}(x)$. Then, based on the orthogonality of $|\Phi_{j\sigma}\rangle$, the spin-dependent tunnelings can be calculated as
\begin{align}
t^{}_{\sigma}=&\langle\Phi^{}_{1\sigma}|\Delta V_{2}(x)|\Phi^{}_{2\sigma}\rangle, \nonumber\\
t'_{\sigma}=&\langle\Phi^{}_{1\sigma}|\Delta V_{2}(x)|\Phi^{}_{2\bar{\sigma}}\rangle.
\end{align}

Up to the first order in $\exp(-d^{2}/x^{2}_{0})$, the spin-conserved and spin-flipped tunnelings can be explicitly written as
\begin{align}
&t^{}_{\Downarrow}=t^{\ast}_{\Uparrow}=-3\left(1+\frac{1}{\chi^{2}+1}\frac{x^{2}_{0}}{d^{2}}\right)V_{0} s_{d},  \nonumber \\
&t'_{\Downarrow}=t'_{\Uparrow}=-3\left(1+\frac{1}{\chi^{2}+1}\frac{x^{2}_{0}}{d^{2}}\right)V_{0} s_{ x}.
\end{align}
The presence of the $\chi$-dependent terms in $s_{x}$ and $s_{d}$, cf. Eq.~(\ref{overlaps}), indicates the contributions of the high orbital states to the spin tunnelings. However, because in the system we consider, the magnetic field $B\leq20$~mT in the experiments and the orbital splitting $\hbar\omega\sim6$~meV, so the degree of the orbital hybridization is quite small $\chi \sim 4\times 10^{-3}$. Thus, compared with the orbital ground states, the effect of the high orbital states on the spin tunnelings can be neglected, and the spin tunnelings can be simplified as
\begin{widetext}
\begin{align}
&t^{}_{\Downarrow}=t^{\ast}_{\Uparrow}=-3V_{0}e^{-d^{2}/x^{2}_{0}}\left(1+\frac{x^{2}_{0}}{d^{2}}\right)\left[\cos\left(\frac{2d}{x_{\rm so}}\right)-i\cos\gamma\sin\left(\frac{2d}{x_{\rm so}}\right)\right] \nonumber\\
&t'_{\Downarrow}=t'_{\Uparrow}=-3iV_{0}e^{-d^{2}/x^{2}_{0}}\left(1+\frac{x^{2}_{0}}{d^{2}}\right)\sin\left(\frac{2d}{x_{\rm so}}\right)\sin\gamma.
\label{tuns}
\end{align}
\end{widetext}
Similarly, the single-electron energy is given by
\begin{align}
&\varepsilon^{}_{j\Uparrow/\Downarrow}=\frac{3x^{4}_{0}}{4d^{4}}V_{0}+\frac{1}{2}(\hbar\omega-m_{e}\alpha^{2})\pm\frac{1}{2}g\mu_{\rm B}Bf.
\label{e-o}
\end{align}

\section{\label{Appendix_C}~ Derivation of the anisotropic exchange Hamiltonian}

As given in the main text, when keeping only the leading Coulomb-interaction terms, the second-quantization Hamiltonian describing the Coulomb interaction between two electrons in a nanowire double quantum dot (DQD) reduces to
\begin{align}
H_{\rm c}=&\frac{U^{}}{2}\sum_{j=1}^2\sum_{\sigma}n^{}_{j\sigma}n^{}_{j\bar{\sigma}}
+\frac{U'}{2}\sum_{j\neq j'}\sum_{\sigma\sigma'}n^{}_{j\sigma}n^{}_{j'\sigma'},
\end{align}
where $n^{}_{j\sigma}=c^{\dagger}_{j\sigma}c^{}_{j\sigma}$ is the particle number operator and $U^{}$ ($U'$) represents the intradot (interdot) Coulomb repulsion. When the single-electron Hamiltonian $H_{0}$ in Eq.~(\ref{s-q}) is included, the total Hamiltonian of the two electrons in the nanowire DQD is
\begin{equation}
H=H_{0}+H_{\rm c}.
\end{equation}

In the regime of strong intradot Coulomb repulsion, i.e., $U-U'\gg |t^{}_{\sigma}|,|t'_{\sigma}|$, each QD can only be occupied by one electron, and the Hilbert space we investigate is restricted to the subspace defined by the projection operator
\begin{align}
\mathbf{P}=&\frac{1}{2}\sum_{j\neq j'}\sum_{\sigma \sigma'}n^{}_{j\sigma}n^{}_{j'\sigma'}(1-n^{}_{j\bar{\sigma}})(1-n^{}_{j'\bar{\sigma'}}).
\label{projection}
\end{align}
 Then, the effective Hamiltonian reads
 \begin{align}
 H_{\rm eff}=\mathbf{P}H\mathbf{P}-\mathbf{P}H\mathbf{Q}\frac{1}{\mathbf{Q}H\mathbf{Q}-E}\mathbf{Q}H\mathbf{P},
 \label{sa}
 \end{align}
where the projection operator $\mathbf{P}$ is given in Eq.~(\ref{projection}) and $\mathbf{Q}=\mathbf{1}-\mathbf{P}$, while $E$ is the ground-state energy of the nanowire DQD. In the strong repulsion regime, $\mathbf{Q}H\mathbf{Q}-E$ can be approximated as $U-U'$.  Based on the definition of the
 projection operator, the first term on the right side of Eq.~(\ref{sa}) can be rewritten as
 \begin{align}
 \mathbf{P}H\mathbf{P}=\left(\varepsilon^{}_{1\Uparrow}-\varepsilon^{}_{1\Downarrow}\right)S^{z}_{1}+\left(\varepsilon^{}_{2\Uparrow}-\varepsilon^{}_{2\Downarrow}\right)
 S^{z}_{2} \nonumber \\
 +\frac{\varepsilon^{}_{1\Uparrow}+\varepsilon^{}_{1\Downarrow}}{2}+\frac{\varepsilon^{}_{2\Uparrow}+\varepsilon^{}_{2\Downarrow}}{2}+4U',
 \label{ser2}
 \end{align}
where $\mathbf{S}^{}_{j=1,~2}=(1/2)\sum_{\sigma,\sigma'=\Uparrow,\Downarrow}c^{\dagger}_{j\sigma}\hat{\boldsymbol{\sigma}}^{}_{\sigma\sigma'}c^{}_{ j \sigma'}$ are the pseudo-spin operators, and
the explicit forms of $\varepsilon^{}_{j\sigma}$ ($j=1,2;~\sigma=\Uparrow,\Downarrow$) are given in Eq.~(\ref{e-o}).

In a QD, the electron occupancy can only be affected by the tunneling Hamiltonian $H_{\rm t}$, so $\mathbf{P}H\mathbf{Q}$ can be simplified as $\mathbf{P}H_{\rm t}\mathbf{Q}$. Similarly,
we have $\mathbf{Q}H\mathbf{P}=\mathbf{Q}H_{\rm t}\mathbf{P}$ and $\mathbf{P}H_{\rm t}\mathbf{P}=0$.  Thus, the second term on the right side of Eq.~(\ref{sa}) can be rewritten as
$-\mathbf{P}H^{2}_{\rm t}\mathbf{P}/(U-U')$. Taking the explicit form of $H_{\rm t}$ into consideration, $\mathbf{P}H^{2}_{\rm t}\mathbf{P}$ can be
expanded as
\begin{widetext}
\begin{align}
\mathbf{P}H^{2}_{\rm t}\mathbf{P}=&\sum_{\sigma,\sigma'}[|t^{}_{\Uparrow}|^{2}(\mathbf{P}c^{\dagger}_{1\sigma}c^{}_{1\sigma'}\mathbf{P}\mathbf{P}c^{}_{2\sigma}c^{\dagger}_{2\sigma'}\mathbf{P}
+\mathbf{P}c^{\dagger}_{2\sigma}c^{}_{2\sigma'}\mathbf{P}\mathbf{P}c^{}_{1\sigma}c^{\dagger}_{1\sigma'}\mathbf{P})
+t^{}_{\Downarrow}t'^{\ast}_{\Uparrow}(\mathbf{P}c^{\dagger}_{1\sigma}c^{}_{1\sigma'}\mathbf{P}\mathbf{P}c^{}_{2\sigma}c^{\dagger}_{2\bar{\sigma'}}\mathbf{P}
+\mathbf{P}c^{\dagger}_{2\sigma}c^{}_{2\sigma'}\mathbf{P}\mathbf{P}c^{}_{1\bar{\sigma}}c^{\dagger}_{1\sigma'}\mathbf{P}) \nonumber \\
&+t^{\ast}_{\Downarrow}t'_{\Uparrow}(\mathbf{P}c^{\dagger}_{1\sigma}c^{}_{1\sigma'}\mathbf{P}\mathbf{P}c^{}_{2\bar{\sigma}}c^{\dagger}_{2\sigma'}\mathbf{P}
+\mathbf{P}c^{\dagger}_{2\sigma}c^{}_{2\sigma'}\mathbf{P}\mathbf{P}c^{}_{1\sigma}c^{\dagger}_{1\bar{\sigma'}}\mathbf{P})+
|t'_{\Uparrow}|^{2}(\mathbf{P}c^{\dagger}_{1\sigma}c^{}_{1\sigma'}\mathbf{P}\mathbf{P}c^{}_{2\bar{\sigma}}c^{\dagger}_{2\bar{\sigma'}}\mathbf{P}
+\mathbf{P}c^{\dagger}_{2\sigma}c^{}_{2\sigma'}\mathbf{P}\mathbf{P}c^{}_{1\bar{\sigma}}c^{\dagger}_{1\bar{\sigma'}}\mathbf{P})]  \nonumber \\
&+(t^{}_{\Uparrow}t^{\ast}_{\Downarrow}-|t^{}_{\Uparrow}|^{2})(c^{\dagger}_{1\Uparrow}c^{}_{1\Downarrow}c^{}_{2\Uparrow}c^{\dagger}_{2\Downarrow}
+c^{\dagger}_{2\Downarrow}c^{}_{2\Uparrow}c^{}_{1\Downarrow}c^{\dagger}_{1\Uparrow})
+(t^{}_{\Downarrow}t^{\ast}_{\Uparrow}-|t^{}_{\Uparrow}|^{2})(c^{\dagger}_{1\Downarrow}c^{}_{1\Uparrow}c^{}_{2\Downarrow}c^{\dagger}_{2\Uparrow}
+c^{\dagger}_{2\Uparrow}c^{}_{2\Downarrow}c^{}_{1\Uparrow}c^{\dagger}_{1\Downarrow})   \nonumber \\
&+(t^{}_{\Uparrow}-t^{}_{\Downarrow})t'^{\ast}_{\Downarrow}(c^{\dagger}_{1\Uparrow}c^{}_{1\Uparrow}c^{}_{2\Uparrow}c^{\dagger}_{2\Downarrow}
+c^{\dagger}_{2\Downarrow}c^{}_{2\Uparrow}c^{}_{1\Uparrow}c^{\dagger}_{1\Uparrow})
+(t^{}_{\Uparrow}-t^{}_{\Downarrow})t'^{\ast}_{\Downarrow}(c^{\dagger}_{1\Uparrow}c^{}_{1\Downarrow}c^{}_{2\Uparrow}c^{\dagger}_{2\Uparrow}
+c^{\dagger}_{2\Uparrow}c^{}_{2\Uparrow}c^{}_{1\Downarrow}c^{\dagger}_{1\Uparrow})   \nonumber \\
&+(t^{\ast}_{\Uparrow}-t^{\ast}_{\Downarrow})t'_{\Uparrow}(c^{\dagger}_{1\Uparrow}c^{}_{1\Uparrow}c^{}_{2\Downarrow}c^{\dagger}_{2\Uparrow}
+c^{\dagger}_{2\Uparrow}c^{}_{2\Downarrow}c^{}_{1\Uparrow}c^{\dagger}_{1\Uparrow})
+(t^{\ast}_{\Uparrow}-t^{\ast}_{\Downarrow})t'_{\Uparrow}(c^{\dagger}_{1\Downarrow}c^{}_{1\Uparrow}c^{}_{2\Uparrow}c^{\dagger}_{2\Uparrow}
+c^{\dagger}_{2\Uparrow}c^{}_{2\Uparrow}c^{}_{1\Uparrow}c^{\dagger}_{1\Downarrow}).
\label{funt}
\end{align}
Using the identity relations~\cite{Nagaosa1999}
\begin{align}
c^{\dagger}_{j\sigma}c^{}_{j\sigma'}=&\frac{1}{2}\delta^{}_{\sigma\sigma'}\left(n^{}_{j\Uparrow}+n^{}_{j\Downarrow}\right)+\mathbf{S}^{}_{j}\cdot\hat{\boldsymbol{\sigma}}^{}_{\sigma'\sigma}, \nonumber\\
c^{}_{j\sigma}c^{\dagger}_{j\sigma'}=&\delta^{}_{\sigma\sigma'}\Big(1-\frac{n^{}_{j\Uparrow}+n^{}_{j\Downarrow}}{2}\Big)
-\mathbf{S}^{}_{j}\cdot\hat{\boldsymbol{\sigma}}^{}_{\sigma\sigma'},
\label{def}
\end{align}
with $\hat{\boldsymbol{\sigma}^{}_{}}=(\sigma^{}_{x},\sigma^{}_{y},\sigma^{}_{z})$, we can rewrite Eq.~(\ref{funt}) as
\begin{align}
\mathbf{P}H^{2}_{\rm t}\mathbf{P}=&2|t^{}_{\Uparrow}|^{2}(\frac{1}{2}-2\mathbf{S}^{}_{1}\cdot\mathbf{S}^{}_{2})+2|t'_{\Uparrow}|^{2}(\frac{1}{2}+2\mathbf{S}^{}_{1}\cdot\mathbf{S}^{}_{2}-4S^{x}_{1}S^{x}_{2})
\nonumber \\
&+4i(t^{}_{\Downarrow}t'^{\ast}_{\Uparrow}-t^{\ast}_{\Downarrow}t'_{\Uparrow})(S^{z}_{1}\cdot S^{y}_{2}-S^{z}_{2}\cdot S^{y}_{1})+W^{}_{1}+W^{}_{2}+W^{}_{3},
\label{ser3}
\end{align}
where
\begin{align}
W^{}_{1}=&(t^{\ast}_{\Uparrow}-t^{\ast}_{\Downarrow})t'_{\Uparrow}(c^{\dagger}_{1\Uparrow}c^{}_{1\Uparrow}c^{}_{2\Downarrow}c^{\dagger}_{2\Uparrow}
+c^{\dagger}_{2\Uparrow}c^{}_{2\Downarrow}c^{}_{1\Uparrow}c^{\dagger}_{1\Uparrow})
+(t^{\ast}_{\Uparrow}-t^{\ast}_{\Downarrow})t'_{\Uparrow}(c^{\dagger}_{1\Downarrow}c^{}_{1\Uparrow}c^{}_{2\Uparrow}c^{\dagger}_{2\Uparrow}
+c^{\dagger}_{2\Uparrow}c^{}_{2\Uparrow}c^{}_{1\Uparrow}c^{\dagger}_{1\Downarrow}), \nonumber \\
W^{}_{2}=&(t^{}_{\Uparrow}-t^{}_{\Downarrow})t'^{\ast}_{\Downarrow}(c^{\dagger}_{1\Uparrow}c^{}_{1\Uparrow}c^{}_{2\Uparrow}c^{\dagger}_{2\Downarrow}
+c^{\dagger}_{2\Downarrow}c^{}_{2\Uparrow}c^{}_{1\Uparrow}c^{\dagger}_{1\Uparrow})
+(t^{}_{\Uparrow}-t^{}_{\Downarrow})t'^{\ast}_{\Downarrow}(c^{\dagger}_{1\Uparrow}c^{}_{1\Downarrow}c^{}_{2\Uparrow}c^{\dagger}_{2\Uparrow}
+c^{\dagger}_{2\Uparrow}c^{}_{2\Uparrow}c^{}_{1\Downarrow}c^{\dagger}_{1\Uparrow}), \label{P-S} \\
W^{}_{3}=&(t^{}_{\Uparrow}t^{\ast}_{\Downarrow}-|t^{}_{\Uparrow}|^{2})(c^{\dagger}_{1\Uparrow}c^{}_{1\Downarrow}c^{}_{2\Uparrow}c^{\dagger}_{2\Downarrow}
+c^{\dagger}_{2\Downarrow}c^{}_{2\Uparrow}c^{}_{1\Downarrow}c^{\dagger}_{1\Uparrow})
+(t^{}_{\Downarrow}t^{\ast}_{\Uparrow}-|t^{}_{\Uparrow}|^{2})(c^{\dagger}_{1\Downarrow}c^{}_{1\Uparrow}c^{}_{2\Downarrow}c^{\dagger}_{2\Uparrow}
+c^{\dagger}_{2\Uparrow}c^{}_{2\Downarrow}c^{}_{1\Uparrow}c^{\dagger}_{1\Downarrow})\nonumber.
\end{align}
With Eq.~(\ref{def}), we have the following relationships:
\begin{align}
c^{\dagger}_{j\Uparrow}c^{}_{j\Downarrow}=&S_{j}^{x}+iS_{j}^{y}, ~~~ c^{\dagger}_{j\Downarrow}c^{}_{j\Uparrow}=S_{j}^{x}-iS_{j}^{y}, \nonumber \\
c^{\dagger}_{j\Downarrow}c^{}_{j\Downarrow}=&\frac{1}{2}-S_{j}^{z}, ~~~ c^{\dagger}_{j\Uparrow}c^{}_{j\Uparrow}=\frac{1}{2}+S_{j}^{z},
\end{align}
which can be used to simplify the terms in Eq.~(\ref{P-S}). Then, we obtain
\begin{align}
W^{}_{1}
=&
-2(t^{\ast}_{\Uparrow}-t^{\ast}_{\Downarrow})t'_{\Downarrow}[S^{z}_{1}(S^{x}_{2}+iS^{y}_{2})+S^{z}_{2}(S^{x}_{1}-iS^{y}_{2})],  \nonumber \\
W^{}_{2}=&-2(t^{}_{\Uparrow}-t^{}_{\Downarrow})t'^{\ast}_{\Downarrow}[S^{z}_{1}(S^{x}_{2}-iS^{y}_{2})+S^{z}_{2}(S^{x}_{1}+iS^{y}_{2})], \\
W^{}_{3}= &
-2(\mathbf{S}^{}_{1}\cdot\mathbf{S}^{}_{2}-S^{z}_{1}S^{z}_{2})(t^{}_{\Uparrow}t^{\ast}_{\Downarrow}
-t^{\ast}_{\Uparrow}t^{}_{\Downarrow}-2|t^{}_{\Uparrow}|^{2}) +2i
(\mathbf{S}^{}_{1}\times\mathbf{S}^{}_{2})_{z}(t^{}_{\Uparrow}t^{\ast}_{\Downarrow}-t^{\ast}_{\Uparrow}t^{}_{\Downarrow}) \nonumber.
\end{align}
Because $t^{}_{\Downarrow}=t^{\ast}_{\Uparrow}$ and $t'_{\Downarrow}=t'_{\Uparrow}$ [cf.~Eq(\ref{tuns})],  $W^{}_{1}+W^{}_{2}$ and $W^{}_{3}$ can be finally simplified as
\begin{align}
&W^{}_{1}+W^{}_{2}=4(t^{}_{\Uparrow}-t^{}_{\Downarrow})t'_{\Downarrow}(S^{z}_{1}S^{x}_{2}+S^{x}_{1}S^{z}_{2}), \nonumber \\
&W^{}_{3}=-2(\mathbf{S}^{}_{1}\cdot\mathbf{S}^{}_{2}-S^{z}_{1}S^{z}_{2})(t^{2}_{\Uparrow}-t^{2}_{\Downarrow}-2|t^{}_{\Uparrow}|^{2}) +2i(\mathbf{S}^{}_{1}\times\mathbf{S}^{}_{2})_{z}(t^{2}_{\Uparrow}-t^{2}_{\Downarrow}).
\label{ser4}
\end{align}
By substituting Eqs.~(\ref{ser2}), (\ref{ser3}) and (\ref{ser4}) into Eq.~(\ref{sa}), the effective Hamiltonian reads
 \begin{align}
H_{\rm eff}=&\mathbf{P}H\mathbf{P}-\mathbf{P}H\mathbf{Q}(\mathbf{Q}H\mathbf{Q}-E)^{-1}\mathbf{Q}H\mathbf{P}  \nonumber \\
=&\Delta^{}_{\rm z}(S^{z}_{1}+S^{z}_{2})+2\frac{t^{2}_{\Uparrow}+t^{2}_{\Downarrow}
-2|t'_{\Downarrow}|^{2}}{U-U'}\mathbf{S}^{}_{1}\cdot\mathbf{S}^{}_{2}
-4i\frac{(t^{}_{\Uparrow}+t^{}_{\Downarrow})t'_{\Downarrow}}{U-U'}(\mathbf{S}^{}_{1}\times\mathbf{S}^{}_{2})_{x}
+8\frac{|t'_{\Uparrow}|^{2}}{U-U'}S^{x}_{1}S^{x}_{2} \nonumber \\
&-2\frac{t^{2}_{\Uparrow}+t^{2}_{\Downarrow}-2|t^{}_{\Uparrow}|^{2}}{U-U'}S^{z}_{1}S^{z}_{2}
-2i\frac{t^{2}_{\Uparrow}-t^{2}_{\Downarrow}}{U-U'}(\mathbf{S}^{}_{1}\times\mathbf{S}^{}_{2})_{z}
-4\frac{(t^{}_{\Uparrow}-t^{}_{\Downarrow})t'_{\Downarrow}}{U-U'}(S^{z}_{1}S^{x}_{2}+S^{x}_{1}S^{z}_{2}).
\label{eham}
\end{align}
When the explicit expressions in Eq.~(\ref{tuns}) are used, we obtain the simplified form of the effective Hamiltonian
\begin{align}
H_{\rm eff}=&\Delta^{}_{\rm z}(S^{z}_{ 1}+S^{z}_{ 2})+J\mathbf{S}^{}_{ 1}\cdot\mathbf{S}^{}_{ 2}+\mathbf{D}\cdot\mathbf{S}^{}_{ 1}\times\mathbf{S}^{}_{ 2} +\mathbf{S}^{}_{ 1}\overleftrightarrow{\Gamma}\mathbf{S}^{}_{2},
\end{align}
which is Eq.~(2) in the main text and the exchange coupling strengths are given by
\begin{equation}
J=J_{0}\cos^{2}\left(2d/x_{\rm so}\right),~~~~
\mathbf{D}=J_{0}\sin\left(4d/x_{\rm so}\right)\hat{v},~~~~
\overleftrightarrow{\Gamma}=J_{0}\sin^{2}\left(2d/x_{\rm so}\right)(2\hat{v}\hat{v}-1),
\label{sparameters}
\end{equation}
with the unit vector
$\hat{v}=\mathbf{e}_{z}\cos\gamma-\mathbf{e}_{x}\sin\gamma$.
\end{widetext}

\begin{figure}
\includegraphics[width=0.42\textwidth]{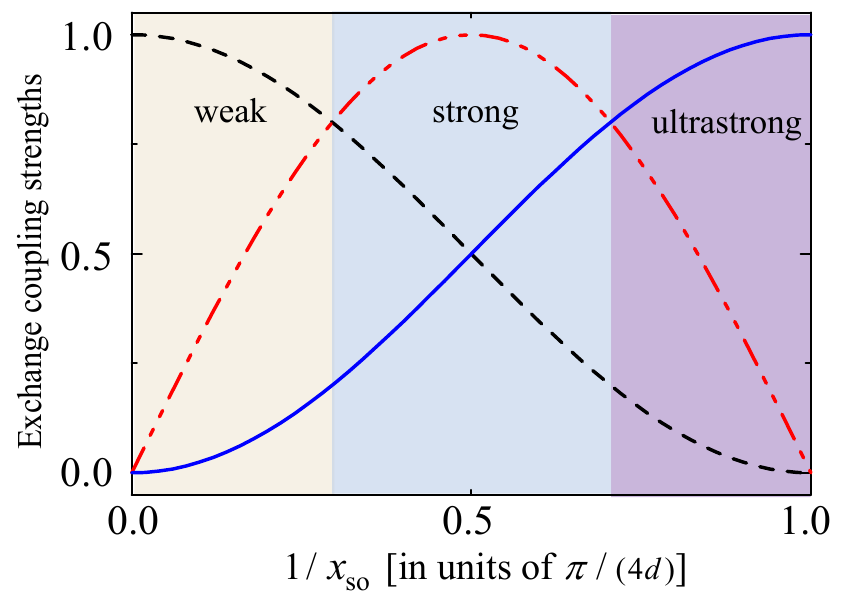}
\caption{(color online)  Exchange-coupling strengths $J$, $|\widehat{D}|$, and
$\|\mathord{\buildrel{\lower3pt\hbox{$\scriptscriptstyle\leftrightarrow $}}\over \Gamma}\|$ (in units of $J_{0}$) as a function of the SOC strength, i.e., $1/x_{\rm so}$. The solid (blue) curve shows the variation of $\|\mathord{\buildrel{\lower3pt\hbox{$\scriptscriptstyle\leftrightarrow $}}\over \Gamma}\|$. The dashed (black) and dot-dashed (red) curves represent the variations of $J$ and $|\widehat{D}|$, respectively. }
\label{Fig2}
\end{figure}

The exchange interaction consists of three parts: the antiferromagnetic $J$ term (i.e., the isotropic Heisenberg interaction), the Dzyaloshinskii-Moriya (DM) interaction ($\widehat{D}$ term), and the symmetric tensor $\overleftrightarrow{\Gamma}$ term (a ferromagnetic term).
It follows from Eq.~(\ref{sparameters}) that
the strengths of the exchange couplings depend on the relative spin-orbit parameter $2d/x_{\rm so}$. Due to the short-range character of the exchange interaction, i.e, $J_{0}\propto \exp(-d^{2}/x^{2}_{0})$, we only consider the case of varying the spin-orbit length $x_{\rm so}$. In Fig.~\ref{Fig2}, the exchange interaction strengths, $J$, $|\widehat{D}|$ and $\|\overleftrightarrow{\Gamma}\|$ $[J_{0}\sin^{2}(2d/x_{\rm so})]$, as a function of the SOC strength, i.e., $1/x_{\rm so}$, are shown.
Experimentally, the Rashba SOC can be tuned by an external electric field.~\cite{Liang2012,Nitta1997} In the present study, we concentrate on the case of a strong SOC. The two-electron exchange interaction is dominated
by the antiferromagnetic interaction in the weak SOC regime, but the DM interaction becomes important in the strong SOC regime. In the ultrastrong SOC regime,  the ferromagnetic term becomes important as well. At present, most of the semiconductor nanowire DQDs are in the weak SOC regime,
but the SOC in an InSb or InAs nanowire DQD nearly falls in the strong-coupling regime,~\cite{Perge2012,Petersson2012,Kouwenhoven2010} which indicates the importance of the anisotropic DM interaction in the spin dynamics.